\documentclass{article}
\usepackage[english]{babel}
\usepackage[tbtags]{amsmath}
\usepackage{amsfonts,amssymb}

\numberwithin{equation}{section}

\newcommand{\mcA}{\mathcal{A}}

\newcommand{\mcP}{\mathcal{P}}
\newcommand{\hC}{\widehat{C}}
\newcommand{\hCp}{\widehat{C}_+}
\newcommand{\hCm}{\widehat{C}_-}
\newcommand{\hc}{\hat{c}}
\newcommand{\bc}{\bar{c}\,}
\newcommand{\bI}{\mathbf{I}}
\newcommand{\bP}{\mathbf{P}}
\newcommand{\bK}{\mathbf{K}}

\newcommand{\mfg}{\mathfrak{g}}
\newcommand{\obI}{\overline{\mathbf{I}}}
\newcommand{\Pe}{\mathcal{P}^{(\epsilon)}}

\DeclareMathOperator{\ad}{ad}
\DeclareMathOperator{\tr}{tr}
\DeclareMathOperator{\Sym}{Sym}
\DeclareMathOperator{\proj}{P}
\DeclareMathOperator{\oproj}{\overline{P}}

\begin{document}
\thispagestyle{empty}
\begin{flushright}
\today\\
\end{flushright}\vspace{3cm}
\begin{center}
{\Large\bf Projectors on invariant subspaces of representations~$\operatorname{ad}^{\otimes 2}$ of Lie~algebras $so(N)$ and $sp(2r)$
and~Vogel parametrization}
\end{center}
\vspace{1cm}

\begin{center}
{\large\bf  A.~P.~Isaev${}^{a,b}$, A.~A.~Provorov$^{a,c}$}
\end{center}

\vspace{0.2cm}

\begin{center}
{${}^a$ \it
Bogoliubov Laboratory of Theoretical Physics,
Joint Institute for Nuclear Research, Dubna, Moscow region,
Russia}\vspace{0.2cm}

${}^b$ {\it
Faculty of Physics,
M.~V.~Lomonosov Moscow State University,
Moscow, Russia}\vspace{0.2cm}

{${}^c$ \it Moscow Institute of Physics and Technology (National Research University), Dolgoprudny, Moscow Region, Russia}
\vspace{0.5cm}

{\tt isaevap@theor.jinr.ru, aleksanderprovorov@gmail.com}
\end{center}
\vspace{2cm}

\begin{abstract}\noindent
Explicit formulae for the projectors onto invariant subspaces in the space of $\ad^{\otimes 2}$ representation of the Lie algebras $so(N)$ and $sp(2r)$ have been found by means of the split Casimir operator. These projectors and characteristic identities
 for the split Casimir operator were considered from the viewpoint of the universal description of simple Lie algebras by using the Vogel parametrisation.
\end{abstract}

\vskip 1.5cm


\newpage
\setcounter{page}{1}
\setcounter{equation}{0}


\section{Introduction}
\label{s1}

In modern theoretical physics the Yang-Baxter equations are
considered to be one of the most important objects to examine. This equation initially appeared in works of J. B. McGuire \cite{1} and C. N. Yang~\cite{2} and plays a key role in the study of  quantum integrable systems~\cite{3},~\cite{4}. In particular, within the quantum inverse scattering method~\cite{5} new structures appeared, which later led to creation of the theory of quantum groups \cite{6},~\cite{7} (see also Refs. therein)
 that describe symmetries of quantum integrable systems (see e.g. \cite{8}--\cite{10}). Note that the Yang-Baxter equations are
 used to formulate quantum groups, as well as to reveal their properties~\cite{11}--\cite{13}. An important class of
 solutions of the Yang-Baxter equations consists
 of those solutions that are invariant under the action of a Lie group $G$ (or its Lie algebra $\mcA$) in a particular representation $T$. Let $V$ be the representation space of~$T$. In this case a solution of the Yang-Baxter equation (the $R$-matrix) is an operator in the representation space $V\otimes V$ of $T$. Assume that the representation $T\otimes T$ is completely reducible and is decomposed into the irreducible representations $T_\lambda$ as follows: $T\otimes T=\sum_\lambda T_\lambda$, where the index $\lambda$ numerates the irreducible representations.
Then the
 required solutions of the Yang-Baxter equation can be
 expanded as a sum
 of the projectors onto the invariant subspaces
  $V_\lambda \subset V \otimes V$ of the representations
  $T_\lambda$ with some coefficients being functions of spectral parameters. In order to find
  explicitly such solutions it is useful to have exact
  expressions for the projectors onto the subspaces $V_\lambda$.

In the case when $T=\ad$ is the adjoint representation, we construct the projectors onto the invariant subspaces of
 $T\otimes T=\ad\otimes \ad$ (our approach is closely related to the approach of \cite{23}, where the diagram technique is used). The construction of the projectors onto the invariant subspaces of the representation $T\otimes T=\ad\otimes \ad$ has one more significance.
 It is related to the notion of \textit{the universal Lie algebra}, which was introduced by P. Vogel in~\cite{14} (see also \cite{15}, \cite{16}; see also historical remarks in \cite{23}, Sect.21.2).
The universal Lie algebra was supposed to be a model of all complex simple Lie algebras $\mcA$. For example, many quantities that characterise an algebra $\mcA$ in different representations $T_\lambda$ (in this case possibly reducible), that participate in the decomposition,
$\ad^{\otimes k}=\sum_{\lambda}T_\lambda$, where $k \ge 1$, can be expressed analytically as functions of the three Vogel parameters (see their definition in Section \ref{s5}). These parameters take specific values for each of the complex simple Lie algebras~$\mcA$ (see e.g. \cite{17}, and Section ~\ref{s5} below). In particular it was shown, that using the Vogel parameters one can express the dimensions of the representations $T_\lambda$ in the cases of $k=2,3$ \cite{14}, the dimension of an arbitrary representation $T_{\lambda'}$ with the highest weight $\lambda'=k\lambda_{\ad}$, where $\lambda_{\ad}$~is the highest root of a Lie algebra~$\mcA$~\cite{18}, as well as values of the higher Casimir operators in the adjoint represenation of a Lie algebra $\mcA$ \cite{17}. Besides, in paper~\cite{19} it was shown, that the universal description of complex simple Lie algebras allows to formulate some types of knot polynomials as a single function
 for all the simple Lie algebras simultaneously.
 In our paper we also demonstrate a particular example of the universal Lie algebra description.
 Namely, the projectors onto invariant subspaces of the tensor product of two adjoint representations of the Lie algebra $so(N,\mathbb{C})$ for $N\ge 3$ and $sp(N,\mathbb{C})$ for
 $N=2r\ge 2$ are written in
a unified form that illustrates
 a correspondence between some structures related to $so(N,\mathbb{C})$ and $sp(N,\mathbb{C})$ Lie algebras. The correspondence is
 given
 by substitution $N\to -N$ (see \cite{23} and references therein). These results having been expressed
 in terms of the Vogel parameters, completely agree with
 consideration in papers \cite{14},~\cite{17}, \cite{18}.

\section{A definition of the algebras $so(N,\mathbb{C})$ and $sp(N,\mathbb{C})$}
\label{s2}
In this Section, to fix notation,
 we give the well known definition of the Lie
 algebras $so(N)$ and $sp(2n)$ which
 one can find in many monographs and textbooks
 (see e.g. \cite{23w}, \cite{23} and \cite{20}). We prefer to give
 a natural unified definition of these algebras
 since it will be usefull for us in next Sections.

In order to describe the Lie algebras $so(N,\mathbb{C})$ and $sp(N,\mathbb{C})$ uniformly, let us introduce the space $V_N$ of their defining representations. The metric $||c_{ij}||_{i,j=1,\dots, N}$ on $V_N$ is defined to be the identity $(N\times N)$ matrix $I_N$ in the case of algebra $so(N,\mathbb{C})$ and the antisymmetric $(N\times N)$ matrix
\begin{equation}
\label{eq2.1}
\|c_{ij}\| =
\begin{pmatrix}
\hphantom{-}0 & I_{r}\\
-I_{r} & 0
\end{pmatrix}
\end{equation}
in the case of algebra $sp(N,\mathbb{C})$, where $N=2r$ is an even number.

We have thus $c_{ij}=\epsilon c_{ji}$, where the parameter $\epsilon$ takes the value $+1$ in the case of algebra~$so(N,\mathbb{C})$, and $-1$ in the case of algebra $sp(N,\mathbb{C})$ (from where we have $\epsilon^2=1$). The inverse matrix $\bc^{ij}$ of the metric $c$ is defined as follows: $\bc^{ik}c_{kj}=\delta^i_j$. With the help of the matrices $c$ and $\bc$ one can raise and lower indices: $z_{i_1}{}^{j_2j_3\dots}=c_{i_1j_1}z^{j_1j_2j_3\dots}$
and~$z^{j_1}{}_{i_2i_3\dots}=\bc^{j_1i_1}z_{i_1i_2i_3\dots}$.

Using the matrix identities $(e_s{}^r)^i{}_k=\delta^r_k\delta^i_s$, which form a basis of the algebra of linear operators on~$V_N$, one can write bases of both $so(N,\mathbb{C})$ and $sp(N,\mathbb{C})$ algebras in the defining representation. Lowering the index $r$ yields $(e_{sr})^i{}_k=c_{rk}\delta^i_s$. In these terms the generators of the Lie algebras $so(N,\mathbb{C})$
and~$sp(N,\mathbb{C})$ can be expressed as follows:
\begin{align}
\label{eq2.2}
&M_{ij}=e_{ij}-\epsilon e_{ji},\\
&(M_{ij})^k{}_l=c_{jl}\delta^k_i-\epsilon c_{il}\delta^k_j=2\delta_{[i}^kc_{j)l},
\label{eq2.3}
\end{align}
where the notation $[ij)$ implies antisymmetrization in the case of algebra~$so(N,\mathbb{C})$ and symmetrization in the case of algebra~$sp(N,\mathbb{C})$. The commutation relation of both algebras are given by the following formula:
\begin{equation}
\label{eq2.4}
[M_{ij},M_{kl}]=c_{jk}M_{il}-\epsilon c_{ik}M_{jl}-\epsilon c_{jl}M_{ik}+c_{il}M_{jk}=X_{ij,kl}{}^{mn}M_{mn},
\end{equation}
from where we can get the structure constants of the Lie algebras under consideration in the basis \eqref{eq2.2}:
\begin{equation}
\label{eq2.5}
X_{ij,kl}{}^{mn}=c_{jk}\delta_{i}^{[m}\delta_{l}^{n)}-\epsilon c_{ik}\delta_{j}^{[m}\delta_{l}^{n)}-\epsilon c_{jl}\delta_{i}^{[m}\delta_{k}^{n)}+c_{il}\delta_{j}^{[m}\delta_{k}^{n)}.
\end{equation}
They can also be rewritten in a more concise manner:
\begin{equation}
\label{eq2.6}
X_{i_1i_2,j_1j_2}{}^{k_1k_2}=4\Sym^\epsilon_{1\leftrightarrow 2}(c_{i_2j_1}\delta_{i_1}^{k_1}\delta_{j_2}^{k_2}),
\end{equation}
where $\Sym^\epsilon_{1\leftrightarrow 2}$ denotes (anti)symmetrization over the following pairs of indices $(i_1,i_2)$, $(j_1,j_2)$, $(k_1,k_2)$. For example,
 $\Sym^\epsilon_{1\leftrightarrow 2}(x_{i_1 i_2}) \equiv
 x_{i_1 i_2} - \epsilon x_{i_2 i_1}$.
Hereinafter we will be using the designation $\mfg_N^\epsilon$ to mean the algebra $so(N,\mathbb{C})$ whenever $\epsilon=+1$ and $sp(N,\mathbb{C})$ if $\epsilon=-1$.

Let us note that (anti)symmetrized pairs of indices $(i_1,i_2)$, $(j_1,j_2)$ and $(k_1,k_2)$, being the indices of the basis vectors $M_{i_1 i_2}$
 of the algebra $\mfg_N^\epsilon$, can also be viewed as coordinate indices in the space of the adjoint representation of the algebra $\mfg_N^\epsilon$.

\section{The split Casimir operator}
\label{s3}

In this section we will describe a general procedure of building the projectors onto invariant subspaces of the representation $\ad^{\otimes 2}(\mcA)$ of a complex simple Lie algebra $\mcA$ by using the algebra's split Casimir operator $\hC$. We will also find explicit formulae for the operator $\hC$ in the representation $\ad^{\otimes 2}(\mcA)$ when $\mcA=so(N,\mathbb{C})$
and~$\mcA=sp(N,\mathbb{C})$.

\subsection{The split Casimir operator in highest weight representations}
\label{ss3.1}

Let $\mathcal{A}$ be a simple Lie algebra with the basis elements $X_a$
and the structure relations
\begin{equation}
\label{eq3.1}
[X_a,X_b]= C^d_{ab}X_d,
\end{equation}
where $C^d_{ab}$~are the structure constants. The Cartan-Killing metric is defined in the standard fashion:
\begin{equation}
\label{eq3.2}
{\sf g}_{ab} \equiv C^{d}_{ac} \, C^{c}_{bd} =
\tr(\ad(X_a)\cdot \ad(X_b)),
\end{equation}
 and the structure constants $C_{abc} \equiv C^{d}_{ab} \, {\sf g}_{dc}$
 are antisymmetric in the indices $(a,b,c)$.
$\mathcal{U}(\mathcal{A})$ denotes the universal enveloping algebra of $\mathcal{A}$. Consider the operator
\begin{equation}
\label{eq3.3}
\hC  = {\sf g}^{ab} X_a \otimes
  X_b \in \mathcal{A}  \otimes  \mathcal{A}
   \subset \mathcal{U}(\mathcal{A}) \otimes\mathcal{U}(\mathcal{A}),
\end{equation}
 where the matrix $\|{\sf g}^{ab}\|$ is the inverse matrix of $\|{\sf g}_{ab}\|$, which, in turn, defines the Cartan-Killing metric \eqref{eq3.2}:
\begin{equation}
\label{eq3.4}
 {\sf g}^{ab}{\sf g}_{bc}=\delta^a_c.
 \end{equation}
 The operator $\hC$ is called the \textit{split} (or \textit{polarised}) \textit{Casimir operator of the Lie algebra} $\mathcal{A}$.
This operator is related to the usual quadratic Casimir operator
\begin{equation}
\label{eq3.5}
 C_{(2)} = {\sf g}^{ab}X_a \cdot X_b
\end{equation}
according to the formula
\begin{equation}
\label{eq3.6}
 \Delta(C_{(2)}) = C_{(2)} \otimes I + I \otimes C_{(2)} + 2 \, \hC ,
\end{equation}
where $\Delta$~is the comultiplication:
\begin{equation}
\label{eq3.7}
\Delta(X_a) = (X_a \otimes I + I \otimes X_a).
\end{equation}

Let $T_1$ and $T_2$~be two irreducible representations the highest weights of which are $\lambda_1$ and $\lambda_2$, and which act on the spaces $\mathcal{V}_1$ and $\mathcal{V}_2$. Then, using the formula \eqref{eq3.6} and the decomposition
\begin{equation}
\label{eq3.8}
 \mathcal{V}_1\otimes \mathcal{V}_2 =
 \sum_\lambda \mathcal{V}_\lambda
\end{equation}
 where $\mathcal{V}_\lambda$
 are the subspaces of the irreducible representations $T_\lambda$ with the highest weights $\lambda$, we can deduce the following relations:
\begin{equation}
\label{eq3.9}
\begin{aligned}
 T(\hC )\cdot  (\mathcal{V}_1\otimes \mathcal{V}_2)={}&
 \frac{1}{2} \sum_\lambda (c_{2}^{(\lambda)} -c_{2}^{(\lambda_1)} -c_{2}^{(\lambda_2)})
  \mathcal{V}_\lambda\quad\Longleftrightarrow\\
&\Longleftrightarrow \quad T(\hC )\cdot \mathcal{V}_\lambda =\frac{1}{2}
 (c_{2}^{(\lambda)} -c_{2}^{(\lambda_1)} -c_{2}^{(\lambda_2)}) \mathcal{V}_\lambda .
\end{aligned}
\end{equation}
 Here we used the following notation
 $T(\hC ):=(T_1 \otimes T_2 ) \hC $ and utilized the equalities\footnote{The first of those equalities is a consequence of the fact, that for each
 $U \in \mathcal{U}(\mathcal{A})$ $(T_1 \otimes T_2 )\Delta(U)$ commutes with the projectors $P_\lambda$,
 which distinguish the irreducible subrepresentations of $(T_1 \otimes T_2 )$
 and hence are $\operatorname{Ad}$-invariant operators.}
 $$
 \Delta(C_{(2)}) \sum_\lambda \mathcal{V}_\lambda =
 \sum_\lambda c_{2}^{(\lambda)} \mathcal{V}_\lambda, \qquad
 ( C_{(2)} \otimes I + I \otimes C_{(2)}) \mathcal{V}_1\otimes \mathcal{V}_2 =
  (c_{2}^{(\lambda_1)} + c_{2}^{(\lambda_2)}) \mathcal{V}_1\otimes \mathcal{V}_2.
 $$
 Here $c_{2}^{(\lambda)}$ is the value of the quadratic Casimir operator in the representation with the highest weight $\lambda$,
\begin{equation}
\label{eq3.10}
c_{2}^{(\lambda)} = (\lambda,\lambda + 2 \, \delta), \quad\quad
\delta  = \sum_{f=1}^r \lambda_{(f)} = \frac{1}{2} \sum_{\alpha >0} \alpha,
\end{equation}
$\lambda_{(f)}$~being the fundamental weights of the rank-$r$ Lie algebra $\mathcal{A}$, $\alpha$~being the roots of $\mathcal{A}$,
and the summation being carried out over the positive roots $(\alpha >0)$. The metric in the root space is given by the matrix~\eqref{eq3.4}.
Note also, that the formula \eqref{eq3.9} can be used to derive the characteristic identity
\begin{equation}
\label{eq3.11}
{\prod_{\lambda}}^{\, \prime} \biggl(T(\hC)-\frac{1}{2}
(c_2^{(\lambda)} -c_2^{(\lambda_1)}
-c_2^{(\lambda_2)})\biggr) = 0,
\end{equation}
where the product ${\prod}^{\, \prime}$ is performed only over those weights~$\lambda$ in~\eqref{eq3.8}, which correspond to different eigenvalues~$c_2^\lambda$.

In the next sections we will find explicit expressions for the split Casimir operator
$T(\hC )=(T_1\otimes T_2)(\hC )$ for orthogonal and symplectic Lie algebras in the case of $T_1$ and $T_2$~being the adjoint representations: $T_1 = T_2 = \ad$.
Then the characteristic identity \eqref{eq3.11} can be rewritten in the form
\begin{equation}
\label{eq3.12}
{\prod_{\lambda}}^{\, \prime}
\biggl(\ad^{\otimes 2}(\hC)-\frac{1}{2}(c_2^{(\lambda)}-2c_2^{(\lambda_{\ad})})\biggr)=0,
\end{equation}
where $\lambda_{\ad}$~is the highest weight of the adjoint representation of~$\mcA$. Then,
using the characteristic identity~\eqref{eq3.12},
we will explicitly build (for orthogonal and symplectic Lie algebras)
the projectors on the invariant subspaces~$\mathcal{V}_\lambda$ of the irreducible representations $T_\lambda$, comprising the decomposition of $\ad^{\otimes 2}$.

\subsection{The split Casimir operator of the algebras $so(N,\mathbb{C})$ and $sp(N,\mathbb{C})$ in the tensor product of adjoint representations}
\label{ss3.2}

Let us show that the components of the split Casimir operator \eqref{eq3.3} in the adjoint representation $\ad$ of the Lie algebras
 $so(N,\mathbb{C})$ and $sp(N,\mathbb{C})$ can be written via the components of the same operator in the defining representation.	

The Cartan-Killing metric of the algebra \eqref{eq2.4} with the structure constants \eqref{eq2.5} is given by the following relations \cite{20}:
\begin{equation}
\label{eq3.13}
{\sf g}_{i_1i_2,j_1j_2}=
X_{i_1i_2,\ell_1\ell_2}{}^{k_1k_2}X_{j_1j_2,k_1k_2}{}^{\ell_1\ell_2} =
2(N-2\epsilon)(c_{i_2j_1}c_{j_2i_1}-\epsilon c_{i_1j_1}c_{j_2i_2}).
\end{equation}
Thus, the inverse Cartan-Killing metric is of the form:
\begin{equation}
\label{eq3.14}
{\sf g}^{i_1i_2,j_1j_2}=\frac{1}{8(N-2\epsilon)}(\epsilon \bc^{i_1j_2}\bc^{i_2j_1}-\bc^{i_1j_1}\bc^{i_2j_2}).
\end{equation}

In what follows we will utilize the following notation for the operator $\hC$ (that was defined in \eqref{eq3.3}) in the adjoint representations of the algebras $so(N,\mathbb{C})$ and $sp(N,\mathbb{C})$: $\ad^{\otimes 2}(\hC) \equiv \hC_{\ad}$. In the basis \eqref{eq2.4} with the structure constants \eqref{eq2.5} we have:
\begin{align}
(\hC_{\ad})^{k_1k_2k_3k_4}_{j_1j_2j_3j_4}&=
{\sf g}^{i_1i_2i_3i_4} \,
X_{i_1i_2,j_1j_2}{}^{k_1k_2}X_{i_3i_4,j_3j_4}{}^{k_3k_4} = \nonumber\\
&=16{\sf g}^{i_1i_2i_3i_4}\Sym^\epsilon_{1\leftrightarrow 2}(c_{i_2j_1}\delta^{k_1}_{i_1}\delta^{k_2}_{j_2})
\Sym^\epsilon_{3\leftrightarrow 4}(c_{i_4j_3}\delta^{k_3}_{i_3}\delta^{k_4}_{j_4}),
\label{eq3.15}
\end{align}
where $\Sym_{\alpha\leftrightarrow\beta}$,
as earlier, denotes the (anti)symmetrizator over the pairs of indices $(i_\alpha,i_\beta)$, $(j_\alpha,j_\beta)$ and $(k_\alpha,k_\beta)$. In these relations those indices, that were marked with the subindices $1$ and $2$ correspond to the first copy of the algebra $\mfg_N^\epsilon$, while those marked with $3$ and $4$ relate to the second copy of $\mfg_N^\epsilon$ (in accordance with the fact that the basis elements of the algebras $so(N,\mathbb{C})$ and $sp(N,\mathbb{C})$ have two indices each and can be embedded into~$V_N\otimes V_N$ as vector spaces).

The split Casimir operator $\hC$ in the defining representation $T_f^{\otimes 2}(\hC)\equiv\hC_f\colon V_N\otimes V_N\to V_N\otimes V_N$ in the basis \eqref{eq2.2} with the structure constants \eqref{eq2.5} has the following form:
\begin{equation}
\label{eq3.16}
\begin{aligned}
(\hC_f)^{k_1k_3}_{j_1j_3}&={\sf g}^{i_1i_2i_3i_4}(M_{i_1i_2})^{k_1}{}_{j_1}(M_{i_3i_4})^{k_3}{}_{j_3}=4{\sf g}^{i_1i_2i_3i_4}(\delta^{k_1}_{[i_1}c_{i_2)j_1})(\delta^{k_3}_{[i_3}c_{i_4)j_3}).
\end{aligned}
\end{equation}
From the expressions \eqref{eq3.15} and \eqref{eq3.16} one can conclude that
\begin{equation}
 \label{eq3.17}
(\hC_{\ad})^{k_1k_2k_3k_4}_{j_1j_2j_3j_4}=4
\Sym^\epsilon_{1\leftrightarrow2,3\leftrightarrow 4}
\left((\hC_f)^{k_1k_3}_{j_1j_3}\delta^{k_2}_{j_2}\delta^{k_4}_{j_4}\right),
\end{equation}
where $\Sym^\epsilon_{1\leftrightarrow 2,3\leftrightarrow 4}$ means (anti)symmetrization over the four pairs of indices: $(j_1,j_2)$, $(k_1,k_2)$, $(j_3,j_4)$, $(k_3,k_4)$. Here the indices are interpreted as the indices in the space $V_N$ of the defining representation of $so(N,\mathbb{C})$  or $sp(N,\mathbb{C})$. Correspondingly, the split Casimir operator acts on the space~$V_N^{\otimes 2}$ in the defining representation, and on the space $V_N^{\otimes 4}$ in the adjoint representation. Here the first and the last pair of spaces $V_N$ in $V_N^{\otimes 4}$ are antisymmetrized in the case of $so(N,\mathbb{C})$ and symmetrized in the case of $sp(N,\mathbb{C})$. Let us introduce the operator of (anti)symmetrization of the $a$'th and $b$'th spaces:
\begin{equation}
\mcP_{ab}^\epsilon=\frac{1}{2}(I-\epsilon P_{ab}),
\label{eq3.18}
\end{equation}
where $a\neq b$, $a,b=1,\dots,4$, and $P_{ab}\colon V_N^{\otimes 4}\to V_N^{\otimes 4}$ is the permutation operator, that acts on the $a$'th and the $b$'th spaces (e.g.
 $P_{13}$ has the components $(P_{13})^{i_1i_2i_3i_4}{}_{j_1j_2j_3j_4}=
 \delta^{i_1}_{j_3}\delta^{i_3}_{j_1}\delta^{i_2}_{j_2}\delta^{i_4}_{j_4}$).
 Then for the space $V_{\ad}$ of the adjoint representation, as well as for $V_{\ad}^{\otimes 2}$ we have
 $$
 V_{\ad} = \mcP^\epsilon_{12} V_N^{\otimes 2}, \qquad
 V_{\ad} \otimes V_{\ad} = \mcP^\epsilon_{12}  \mcP^\epsilon_{34}
 V_N^{\otimes 4}.
 $$
 Moreover, we can express the relation \eqref{eq3.17} between the Casimir operators in the defining and the ajoint representations in the following fashion:
\begin{equation}
\label{iiii}
 {\hC_{\ad}} = 4\; \mcP_{12}^\epsilon\mcP_{34}^\epsilon(\hC_f)_{13}
\mcP_{12}^\epsilon \mcP_{34}^\epsilon .
\end{equation}
From \eqref{eq3.16} one can derive the following expression for the split Casimir operator in the defining representation (see e.g. \cite{20})
\begin{equation}
(\hC_f)_{13}=\frac{1}{2(N-2\epsilon)}(P_{13}-\epsilon K_{13}) ,
\end{equation}
which is to be inserted into \eqref{iiii}. Here the operators $(\hC_f)_{13}$, $K_{13}$ and
$P_{13}$ act nontrivially only on the first and third spaces $V_N$ in $V_N^{\otimes 4}$, the operator $K_{13}$ has the following components: $K^{i_1i_2i_3i_4}{}_{j_1j_2j_3j_4}=
\bc^{i_1i_3}c_{j_1j_3}\delta^{i_2}_{j_2}\delta^{i_4}_{j_4}$,
and the operator $P_{13}$, as it has already been noted, permutes the first and the third spaces in~$V_N^{\otimes 4}$. Finally we have for $\hC_{\ad}$:
\begin{equation}
\label{3.21}
\hC_{\ad}=\frac{2}{N-2\epsilon} \, \mcP_{12}^\epsilon\mcP_{34}^\epsilon
\; (P_{13}-\epsilon K_{13})\; \mcP_{12}^\epsilon\mcP_{34}^\epsilon.
\end{equation}

\section{The decomposition of $\ad^{\otimes 2}(\mfg_N^\epsilon)$ into the irreducible representations}
\label{s4}

In this section we will use the split Casimir operator in the adjoint representation to build projectors onto invariant subspaces  of the representation $\ad^{\otimes 2}(\mfg_N^\epsilon)$. Taking into account existence of the `arbitrary' isomorphisms $so(3)\cong sl(2)\cong sp(2)$, $so(4)\cong sl(2)+sl(2)$, $so(5)\cong sp(4)$, $so(6)\cong sl(4)$, we will consider the cases of the aforementioned algebras, apart from $so(5)\cong sp(4)$, at the end of the chapter (the algebras $so(5)\cong sp(4)$ fit into the general picture and do not require a separate examination).

\subsection{The characteristic identity for the operator~$\hC$ in the adjoint representation of the algebras $so(N,\mathbb{C})$ and $sp(N,\mathbb{C})$}
\label{ss4.1}
Let us define the spaces $V_{\ad}^\epsilon$ of the adjoint representation of the algebras $so(N)$ and $sp(N)$ as $V_{\ad}^\epsilon=\Pe_{12}V_N^{\otimes 2}$. The algebra $\mfg_N^\epsilon$ coincides with $V_{\ad}^\epsilon$ as a vector space.
We introduce the following operators that act in the space $V_{\ad}^\epsilon\otimes V_{\ad}^\epsilon\subset V_N^{\otimes 4}$:
\begin{equation}
\label{eq4.1}
\begin{aligned}
&\bI=\Pe_{12}\Pe_{34}\equiv\mcP_{12,34}^{(\epsilon)}, \qquad
\bP=\mcP_{12,34}^{(\epsilon)}P_{13}P_{24}\mcP_{12,34}^{(\epsilon)},\\
&\bK=\mcP_{12,34}^{(\epsilon)}K_{13}K_{24}
\mcP_{12,34}^{(\epsilon)},
\end{aligned}
\end{equation}
for which the following relations hold:
\begin{equation}
\label{eq4.2}
\begin{aligned}
&\bI=\bI P_{12}P_{34}=P_{12}P_{34}\bI,\qquad\qquad \bP=P_{12}P_{34}\bI=\bI P_{13}P_{24},\\
&\bP^2=\bI,\qquad\quad  \bK\bP=\bP\bK=\bK,\qquad\quad  \bK^2=\frac{N(N-\epsilon)}{2}\bK
\end{aligned}
\end{equation}
and
\begin{equation}
\label{eq4.3}
\hC_{\ad}\bP=\bP\hC_{\ad},\qquad\qquad\qquad \hC_{\ad}\bK=\bK\hC_{\ad}=-\bK.
\end{equation}
The operators \eqref{eq4.1} are obviously $\ad$-invariant
with respect to the adjoint action of the algebra $\mfg_N^{\epsilon}$.
Hereinafter we will use the following notation $M\equiv \epsilon N$, which allows the last relation in~\eqref{eq4.2} to be rewritten in the form:
\begin{equation}
\label{eq4.4}
\bK^2=\frac{M(M-1)}{2}\bK.
\end{equation}
In order to find the characteristic identity,
it is convenient to introduce the symmetric~$\hCp$ and the antisymmetric $\hCm$ projectors of the operator $\hC_{\ad}$:
\begin{equation}
\label{eq4.5}
\hC_{\pm}=\frac{1}{2}(\bI\pm\bP)\hC_{\ad},
\end{equation}
which satisfy the following relations:
\begin{align}
\label{eq4.6}
\hC_{\pm}\hC_{\mp}=0, \qquad \bP\hC_{\pm}=\pm \hC_\pm, \qquad \bK\hC_+=\hC_+\bK=-\bK.
\end{align}
Substitution of \eqref{3.21} into ~\eqref{eq4.5} yields explicit formulae for the antisymmetric and the symmetric parts of the operator $\hC_{\ad}$:
\begin{align}
&(\hCm)_{12,34}=\frac{1}{N-2\epsilon}\Pe_{12,34}(P_{24}-\epsilon)K_{13}\Pe_{12,34},
\label{eq4.7}\\*
&(\hCp)_{12,34}=\frac{1}{N-2\epsilon}\Pe_{12,34}(2P_{24}-P_{24}K_{13}-\epsilon K_{13})\Pe_{12,34}.
\label{eq4.8}
\end{align}

The characteristic identity for $\hCm$ can be found by direct calculations with the use of the formula \eqref{eq4.7}:
\begin{equation}\label{eq4.9}
\hC_-^2=-\frac{1}{2}\hC_-\quad\iff\quad \hC_-\biggl(\hC_-+\frac{1}{2}\biggr)=0.
\end{equation}
Note also a useful consequence of the relation~\eqref{eq4.9}:
\begin{equation}
\label{eq4.10}
\hCm^k=\biggl(-\frac{1}{2}\biggr)^{k-1}\hCm,\qquad k\ge 1.
\end{equation}

Analogously, one can use the explicit formula \eqref{eq4.8} for $\hCp$, to find an expression for $\hC_+^2$:
\begin{equation}
\label{eq4.11}
\hCp^2=\frac{1}{(M-2)^2}(\bI+\bP+\bK)-\frac{1}{M-2}\,\hCp+\frac{M-8}{2(M-2)^2}\,\Pe_{12,34}K_{13}(1+\epsilon P_{24})\Pe_{12,34}.
\end{equation}
If $M=8$, then the last summand in ~\eqref{eq4.11}
equals zero and the identity for $\hCp$ takes the following form\footnote{The universal form
(for all exceptional Lie algebras ${\cal A}$) of this
relation was obtained in \cite{23}, eq. (17.10),
 in the assumption that there exists no primitive quartic
 ${\cal A}$-invariant. Our definition of $\hCp$
 differs from definition of ${\bf Q}$ in \cite{23}
 by the sign: $\hCp \to - {\bf Q}$.}:
\begin{equation}
\label{eq4.12}
\hC_+^2=-\frac{1}{6}\hC_++\frac{1}{36}(\bI+\bP+\bK).
\end{equation}
If $M\neq 8$, then by multiplying \eqref{eq4.11} by $\hCp$, we arrive at an identity of degree three:
\begin{equation}
\label{eq4.13}
\hC_+^3=-\frac{1}{2}\,\hC_+^2-\frac{M-8}{2(M-2)^2}\,\hC_++\frac{M-4}{2(M-2)^3}(\bI+\bP-2\bK).
\end{equation}
Since for $M=8$ the identity \eqref{eq4.12}
for $\hCp$ has the degree $2$ rather than $3$, the case
$M=8$ is exceptional and will be considered later in the subsection~\ref{ss4.2}.
Note also, that when $M=4$ the last summand in~ \eqref{eq4.13} vanishes, hence the identity \eqref{eq4.13} in the case of algebra $so(4,\mathbb{C})$ is characteristic for the operator $\hCp$ and has the following explicit form:
\begin{equation}
\label{eq4.14}
\hCp^3=-\frac{1}{2}\hCp^2+\frac{1}{2}\hCp.
\end{equation}

For the purpose of getting characteristic identities for $\hCp$ when $M\neq 4,8$ we need to get rid of the operators $\bP$ and $\bK$ in the formula \eqref{eq4.13}. Multiplying it by $C_+$ and using \eqref{eq4.6}, we can express $\bK$ in terms of $\hCp$:
\begin{equation}
\label{eq4.15}
\frac{M-4}{(M-2)^3}\bK=\hC_{+}^4+\frac{1}{2}\hC_{+}^3+\frac{M-8}{2(M-2)^2}\hC_{+}^2-\frac{M-4}{(M-2)^3}\hC_{+}.
\end{equation}
We then substitute $\bK$ from \eqref{eq4.15} into~\eqref{eq4.13} and find the following identity (which also holds in the case $M=4$ as one can check by direct calculations):
\begin{align}
\label{eq4.16}
\hC_{+}^4+\frac{3}{2}\hC_{+}^3+\frac{(M+1)(M-4)}{2(M-2)^2}\hC_{+}^2&+\frac{M^2-12M+24}{2(M-2)^3}\hC_{+}
-\frac{M-4}{2(M-2)^3}(\bI+\bP)=0.
\end{align}
Multiplying one more time \eqref{eq4.16} by $\hC_{+}$ we achieve the following characteristic identity for $\hC_{+}$:
\begin{align}
\label{eq4.17}
\hC_{+}^5+\frac{3}{2}\,\hC_{+}^4+\frac{(M+1)(M-4)}{2(M-2)^2}\,\hC_{+}^3&+\frac{M^2-12M+24}{2(M-2)^3}\,\hC_{+}^2
-\frac{M-4}{(M-2)^3}\,\hC_{+}=0,
\end{align}
which could be also rewritten in the form:
\begin{align}
\label{eq4.18}
\hC_{+}^5=-\frac{3}{2}\,\hC_{+}^4-\frac{(M+1)(M-4)}{2(M-2)^2}\,\hC_{+}^3&
-\frac{M^2-12M+24}{2(M-2)^3}\,\hC_{+}^2
+\frac{M-4}{(M-2)^3}\,\hC_{+}.
\end{align}
For the upcoming calculations we will also need an expression for $\hC_+^6$, which could be gained by multiplying the identity~\eqref{eq4.18} by $\hC_+$ followed by substitution of the known polynomial for $\hC_+^5$ from \eqref{eq4.18}:
\begin{align}
\label{eq4.19}
\hC_{+}^6={}&\frac{7M^2-30M+44}{4(M-2)^2}\,\hC_{+}^4
+\frac{3M^3-17M^2+30M-24}{4(M-2)^3}\,\hC_{+}^3+{}\nonumber\\
&+\frac{3M^2-32M+56}{4(M-2)^3}\,\hC_{+}^2-\frac{3(M-4)}{2(M-2)^3}\,\hC_{+}.
\end{align}

Now we will use the acquired expressions to find the characteristic identity for the split Casimir operator $\hC_{\ad}=\hC_++\hC_-$. We will be looking for such an expression in the form of a $6$-degree polynomial of $\hC_{\ad}$ with the coefficients~$\alpha_i$:
\begin{equation}
\label{eq4.20}
\hC_{\ad}^6+\alpha_5\hC_{\ad}^5+\alpha_4\hC_{\ad}^4
+\alpha_3\hC_{\ad}^3+\alpha_2\hC_{\ad}^2+\alpha_1\hC_{\ad}+\alpha_0.
\end{equation}
We need to find such $\alpha_i$, for which the expression~\eqref{eq4.20} vanishes. From the formula~\eqref{eq4.6} it is clear that $\hC^k=\hC_+^k+\hC_-^k$,
hence vanishing of the polynomial \eqref{eq4.20} yields the equation
\begin{align*}
\hC_{+}^6&+\alpha_5\hC_{+}^5+\alpha_4\hC_{+}^4+\alpha_3\hC_{+}^3
+\alpha_2\hC_{+}^2+\alpha_1\hC_{+}+{}\nonumber\\
&+\hC_{-}^6+\alpha_5\hC_{-}^5+\alpha_4\hC_{-}^4
+\alpha_3\hC_{-}^3+\alpha_2\hC_{-}^2+\alpha_1\hC_{-}+\alpha_0=0.
\end{align*}
Substituting then the expressions for $\hC_{+}^{5,6}$ in terms of $\hC_{+}^{4,3,2,1}$ according to the formulae \eqref{eq4.18}, \eqref{eq4.19} and the expressions for $\hC_{-}^{6,5,4,3,2}$ in terms of $\hC_{-}$ by the formula \eqref{eq4.10} and setting the coefficients of those operators to zero, we get the values of $\alpha_i$:
\begin{alignat*}{2}
&\alpha_0=0, &\qquad &\alpha_1=-\frac{M-4}{2(M-2)^3},\\
&\alpha_2=\frac{M^2-16M+40}{4(M-2)^3}, &\qquad &\alpha_3=\frac{M^3-3M^2-22M+56}{4(M-2)^3},\\
&\alpha_4=\frac{5M^2-18M+4}{4(M-2)^2}, &\qquad &\alpha_5=2.
\end{alignat*}
The characteristic identity for $\hC_{\ad}$ thus takes the form:
\begin{align}
\label{eq4.21}
\hC_{\ad}^6+2\hC_{\ad}^5&+\frac{5M^2-18M+4}{4(M-2)^2}\hC_{\ad}^4+\frac{M^3-3M^2-22M+56}{4(M-2)^3}\hC_{\ad}^3+{}\nonumber\\*
&+\frac{M^2-16M+40}{4(M-2)^3}\hC_{\ad}^2-\frac{M-4}{2(M-2)^3}\hC_{\ad}=0.
\end{align}
The roots of the equation \eqref{eq4.21} can be found explicitly:
\begin{equation}
\label{eq4.22}
\begin{aligned}
&a_1=0,\qquad a_2=-\frac{1}{2},\qquad a_3=-1,\qquad a_4=\frac{1}{M-2},\\
&a_5=-\frac{2}{M-2},\qquad a_6=\frac{4-M}{2(M-2)}.
\end{aligned}
\end{equation}

Let us note that if $M$ takes one of the following values, some of the roots are degenerate (we discard here the values $M=0,1,2$ as they do not correspond to semisimple Lie algebras):
\begin{align}
&M=4 \quad\implies\quad a_1=a_6=0,\quad a_3=a_5=-1,\label{eq4.23}\\
&M=6 \quad\implies\quad a_2=a_5=-\frac{1}{2},\label{eq4.24}\\
&M=8 \quad\implies\quad a_5=a_6=-\frac{1}{3},\label{eq4.25}
\end{align}
from where we see that if $M=4,6,8$ then the characteristic polynomials do not have the degree $6$, as it is in the general case, but the degrees $4,5,5$ respectively (this also manifests itself in the differences between the identities \eqref{eq4.16} and \eqref{eq4.13} when $M=4$ and the identities \eqref{eq4.12}
and~\eqref{eq4.13} when $M=8$). The cases $M=4,6$ will be considered in more details at the end of this section,
while the subsection~\ref{ss4.2}, as it was pointed out before, will be devoted to the case $M=8$.

Taking into account the roots \eqref{eq4.22} of the polynomial in the left hand side of \eqref{eq4.21}, the characteristic identity \eqref{eq4.21} can be factorized:
\begin{equation}
\label{eq4.26}
\hC_{\ad}\biggl(\hC_{\ad}+\frac{1}{2}\biggr)(\hC_{\ad}+1)\biggl(\hC_{\ad}-\frac{1}{M\kern-1pt-2}\biggr)
\biggl(\hC_{\ad}+\frac{2}{M\kern-1pt-2}\biggr)\biggl(\hC_{\ad}+\frac{M\kern-1pt-4}{2(M\kern-1pt-2)}\biggr)=0.
\end{equation}
The form \eqref{eq4.26} of the characteristic identity allows for construction of projectors onto invariant subspaces in~$V_{\ad}\otimes V_{\ad}$. Those projectors are given by the following standard formula
(see e.g. Sect.~3.5 in \cite{23} and Sect.~4.6.4 in \cite{20a}):
\begin{equation}
\label{eq4.27}
\proj_j\equiv \proj_{a_j}=\prod_{\substack{i=1,\\i\neq j\hfill}}^6\frac{\hC-a_i\bI}{a_j-a_i},
\end{equation}
where $a_i$~are the roots \eqref{eq4.22} of the characteristic equation~\eqref{eq4.26}. Using the identities~\eqref{eq4.6}, as well as the formulas~\eqref{eq4.10}, \eqref{eq4.13}, \eqref{eq4.16}, \eqref{eq4.18}, we arrive at the following expressions for the projectors \eqref{eq4.27} in terms of the operators $\bI$, $\bP$, $\bK$, $\hC_+$, $\hC_-$
 (cf. projectors in \cite{23}, Table 10.3):
\begin{equation}
\label{eq4.28}
\begin{aligned}
&\proj_1=\frac{1}{2}(\bI-\bP)+2\hC_{-},  \\
&\proj_2=-2\hC_{-}, \\
&\proj_3=\frac{2\bK}{(M-1)M}\,,  \\
&\proj_4=\frac{2}{3}(M-2)\hC_{+}^2+\frac{M}{3}\,\hC_{+}
+\frac{(M-4)(\bI+\bP)}{3(M-2)}-\frac{2(M-4)\bK}{3(M-2)(M-1)}\,, \\
&\proj_5=-\frac{2(M-2)^2}{3(M-8)}\,\hC_{+}^2-\frac{(M-2)(M-6)}{3(M-8)}\,\hC_{+}
+\frac{(M-4)(\bI+\bP)}{6(M-8)}+\frac{2\bK}{3(M-8)}\,,\\
&\proj_6=\frac{4(M-2)}{M-8}\,\hC_{+}^2+\frac{4}{M-8}\,\hC_{+}
-\frac{4(\bI+\bP)}{(M-2)(M-8)}-\frac{8(M-4)\bK}{M(M-2)(M-8)}\,.
\end{aligned}
\end{equation}
\clearpage
Note that if $M=8$ then the projectors $\proj_5$ and $\proj_6$ are not formally defined. It is a consequence of the fact, that there is a factor $a_5-a_6={(M-8)}/{2(M-2)}$ in the denominators of those projectors in \eqref{eq4.27}. Nevertheless, if one substitutes the formula \eqref{eq4.11} for $\hCp^2$ into the expressions for $\proj_5$ and~$\proj_6$, then the pole at $M=8$ reduces. As a result, we have for $\proj_5$ the following expression
\begin{equation}
\label{eq4.29}
\proj_5=\frac{1}{6}(1-\epsilon(P_{14}+P_{23}+P_{13}+P_{24})+P_{13}P_{24})\Pe_{12,34},
\end{equation}
that does not depend on $M$ and coincides with the complete antisymmetrizer on $V_N^{\otimes 4}$ in the case of algebras $so(N,\mathbb{C})$ ($\epsilon=+1$) or with the complete symmetrizer on $V_N^{\otimes 4}$ in the case of algebras $sp(N,\mathbb{C})$ ($\epsilon=-1$). For the projector~$\proj_6$ we have:
\begin{equation}
\label{eq4.30}
\proj_6=\frac{4}{M-2}\,\Pe_{12,34}K_{13}\biggl[\frac{1}{2}(1+\epsilon P_{24})-\frac{1}{M}~K_{24}\biggr]\Pe_{12,34}.
\end{equation}
Both operators $\proj_5$ and $\proj_6$ are well-defined at $M=8$. However, due to the identity \eqref{eq4.12} being different in comparison to \eqref{eq4.13}, the case of algebra $so(8,\mathbb{C})$ will be considered separately (see Subsection~\ref{ss4.2}).

The dimensions of the eigenspaces $V_{a_i}$ of $\hC_{\ad}$ in the space ${V_{\ad}\otimes V_{\ad}}$ equal to the traces of the corresponding projectors. In order to find them we will firstly compute the following auxiliary traces:
\begin{equation}
\label{eq4.31}
\begin{alignedat}{3}
&\tr \bI=\frac{M^2}{4}(M-1)^2, &\qquad &\tr \bP=\frac{M}{2}(M-1), &\qquad &\tr \bK=\frac{M}{2}(M-1),\\
&\tr \hC_+=\frac{M}{4}(M-1), &\qquad &\tr \hC_+^2=\frac{3M}{8}(M-1),  &\qquad&\tr \hC_-=-\frac{M}{4}(M-1),
\end{alignedat}
\end{equation}
from where we have (see Table 10.3 in \cite{23})
\begin{equation}
\label{eq4.32}
\begin{aligned}
&\dim(V_{a_1})=\tr \proj_1=\frac{1}{8}M(M-1)(M+2)(M-3),\\
&\dim(V_{a_2})=\tr \proj_2=\frac{1}{2}M(M-1),\\
&\dim(V_{a_3})=\tr \proj_3=1,\\
&\dim(V_{a_4})=\tr \proj_4=\frac{1}{12}M(M+1)(M+2)(M-3),\\
&\dim(V_{a_5})=\tr \proj_5=\frac{1}{24}M(M-1)(M-2)(M-3),\\
&\dim(V_{a_6})=\tr \proj_6=\frac{1}{2}(M-1)(M+2).
\end{aligned}
\end{equation}
Note that the sum of those traces
\begin{equation}
\label{eq4.33}
\sum_{i=1}^6\tr\proj_i=\frac{M^2}{4}(M-1)^2=\tr \bI
\end{equation}
coincides with the dimension of $\mfg_N^\epsilon\otimes \mfg_N^\epsilon$, as it should be.

If $M=4,6$, the characteristic identities have degrees $4$ and $5$ respectively, as it was noted in the discussion after \eqref{eq4.13}, as well as after the formulae \eqref{eq4.23}--\eqref{eq4.25}.
 However, all the operators \eqref{eq4.28} are well-defined when $M=4,6$, constructed from the $\ad$-invariant operators $\bI$, $\bP$, $\bK$, $\hCp$, $\hCp^2$, $\hCm$ and satisfy
 by construction the following relations:
\begin{equation}
\label{eq4.34}
\proj_i\proj_j=\delta_{ij}\proj_i,\qquad \sum_{i=1}^6\proj_i=\bI,\qquad i,j=1,\dots, 6 \; .
\end{equation}
Thus, they form a full system of projectors on invariant subspaces of the representation $\ad^{\otimes 2}(so(4))$ and $\ad^{\otimes 2}(so(6))$. A decomposition of the representation $\ad^{\otimes 2}(so(4))$ into irreducible subrepresentations can be easily found, if one uses the isomorphism $so(4)\cong sl(2)+sl(2)$ and the known decomposition $\ad^{\otimes 2}(sl(2))=1+3+5$, while for a decompsotion of $\ad^{\otimes 2}(so(6))$ one can use, for example, the program
 \texttt{LieART}~\cite{21} (see also Table 10.3 in \cite{23}):
\begin{align}
&\ad^{\otimes 2}(so(4))\equiv 6\otimes 6=1+1+3+3+5+5+9+9,
\label{eq4.35}\\
&\ad^{\otimes 2}(so(6))\equiv 15\otimes 15=1+15+15+20+45+45'+84.
\label{eq4.36}
\end{align}
Comparing dimensions of the representations \eqref{eq4.35} and dimensions of the invariant subspaces \eqref{eq4.32} we see that in the case of algebra $so(4,\mathbb{C})$ out of all the projectors \eqref{eq4.28} only $\proj_1$, $\proj_6$, $\proj_3$ and $\proj_5$ are primitive, the first two of which project onto $10$-dimensional, and the second two project onto $1$-dimensional subspaces. The operator $\proj_2$ projects onto the sum of two $3$-dimensional subspaces, while the operator $\proj_4$~projects onto a sum of two $5$-dimensional subspaces. In the case of algebra $so(6,\mathbb{C})$ comparison of the formulae \eqref{eq4.36} and \eqref{eq4.32} shows, that all the projectors apart from $\proj_1$ are primitive, while $\proj_1$ projects onto the sum of two invariant subspaces, each of which is $45$-dimensional.

Let us consider here also the case of algebra $so(3,\mathbb{C})\cong sl(2,\mathbb{C})\cong sp(2,\mathbb{C})$, that is when $M=3$ (the algebra $so(3,\mathbb{C})$) or $M=-2$ (the algebra $sp(2,\mathbb{C})$). Despite all the roots \eqref{eq4.22} of the characteristic polynomial on the left hand side of~\eqref{eq4.21} being different, the identity~\eqref{eq4.26} is not actually characteristic for the operator~$\hC_{\ad}$ when $M=3$ and $M=-2$. This is due to the fact that if $M=3$ then the projectors $\proj_1$, $\proj_4$ and $\proj_5$ are degenerate, while if $M=-2$ then the projectors $\proj_1$, $\proj_4$ and $\proj_6$ nullify (this happens because the quantities from \eqref{eq4.22}, that correspond to the nullifying projectors, are not eigenvalues of $\hC_{\ad}$ in the considered representation). The remaining three projectors form a full system of projectors and, as it is seen from the decomposition
\begin{equation}
\label{eq4.37}
\ad^{\otimes 2}(sl(2))\equiv 3\otimes 3=1+3+5,
\end{equation}
are primitive. Traces of those projectors in the case $M=3$ and $M=-2$, equal to dimensions of the corresponding invariant subspaces, and are in agreement with the isomorphisms $so(3,\mathbb{C})\cong sl(2,\mathbb{C})\cong sp(2,\mathbb{C})$.

For illustration, in Tables \ref{tab1}, \ref{tab2} we provide dimensions \eqref{eq4.32} of irreducible representations in the decomposition of $\ad\otimes \ad$ for some of the simple Lie algebras $so(N)$, $sp(N)$ and for several values $N$. The dimensions in those tables are in accordance with the data
 presented in papers~\cite{23,22}. Note also, that the characteristic identites \eqref{eq4.26} and the dimensions~\eqref{eq4.32} for~$so(N,\mathbb{C})$ and~$sp(N,\mathbb{C})$ turn into each other if one substitutes $N\to -N$. This manifests a duality between the algebras $so(N)$ and $sp(N)$ (see \cite{23} for more details).

\begin{table}[h]

\small

\tabcolsep=1.5em
\centering

\renewcommand{\arraystretch}{1.18}

\vspace*{1mm}
\caption{Dimensions of irreducible representations for algebra~$so(N)$.}
\vspace*{1mm}
\begin{tabular}{|c|c|c|c|c|c|c|}
\hline
$N$ & $\dim_1$ & $\dim_2$ & $\dim_3$ & $\dim_4$ & $\dim_5$ & $\dim_6$\\
\hline
$5$ & $35$ & $10$ & $1$ & $35$ & $5$ & $14$\\
\hline
$7$ & $189$ & $21$ & $1$ & $168$ & $35$ & $27$\\
\hline
$9$ & $594$ & $36$ & $1$ & $495$ & $126$ & $44$\\
\hline
$10$ & $945$ & $45$ & $1$ & $770$ & $210$ & $54$\\
\hline
$11$ & $1434$ & $55$ & $1$ & $1144$ & $330$ & $65$\\
\hline
\end{tabular}

\label{tab1}

\centering
\renewcommand{\arraystretch}{1.18}

\caption{Dimensions of irreducible representations for algebra~$sp(N)$.}
\vspace*{1mm}
\begin{tabular}{|c|c|c|c|c|c|c|}
\hline
$N$ & $\dim_1$ & $\dim_2$ & $\dim_3$ & $\dim_4$ & $\dim_5$ & $\dim_6$\\
\hline
$4$ & $35$ & $10$ & $1$ & $14$ & $35$ & $5$\\
\hline
$6$ & $189$ & $21$ & $1$ & $90$ & $126$ & $14$\\
\hline
$8$ & $594$ & $36$ & $1$ & $308$ & $330$ & $27$\\
\hline
$10$ & $1430$ & $55$ & $1$ & $780$ & $715$ & $44$\\
\hline
$12$ & $2925$ & $78$ & $1$ & $1650$ & $1365$ & $65$\\
\hline
\end{tabular}

\label{tab2}

\end{table}

\subsection{The case of algebra $so(8)$\label{so8}}
\label{ss4.2}
In order to deduce the characteristic identity
 for $\hC_{\ad}$ in the case of algebra $so(8)$ we use the identities \eqref{eq4.9} and \eqref{eq4.12} for the operators $\hC_-$ and $\hC_+$:
\begin{equation}
\label{eq4.38}
\hC_-^2=-\frac{1}{2}\,\hC_- ,\qquad \hC_+^2=-\frac{1}{6}\,\hC_+
 +\frac{1}{36}(\bI+\bP+\bK).
\end{equation}
Multiplying the second relation by $\hC_+$, multiplying then the obtained identity by $(\hC_+ +1)$ and using the condition
 ${\bf K} (\hC_+ +1) =0$ yields the following formula:
 $$
 \hC_{+}(\hC_{+}+1)\biggl(\hC_{+}+\frac{1}{3}\biggr)\biggl(\hC_{+}-\frac{1}{6}\biggr)=0.
 $$
Note that the characteristic identity for the split Casimir operator $\hC_{\ad}=\hC_+ + \hC_-$ can be built analogously to the general case $so(M)$ by introducing indeterminate coefficients:
\begin{equation}
\label{eq4.39}
\hC_{\ad}\biggl(\hC_{\ad}+\frac{1}{2}\biggr) (\hC_{\ad}+1)
\biggl(\hC_{\ad}+\frac{1}{3}\biggr)\biggl(\hC_{\ad}-\frac{1}{6}\biggr)=0 .
\end{equation}
In this way, the operator $\hC_{\ad}$ has the following eigenvalues
$a_1=0$, $a_2 =-1/2$, $a_3=-1$, $a_4 =-1/3$ and $a_5 =1/6$.
Let us write all the projectors onto the invariant subspaces of $\hC_{\ad}$,
that can be derived from \eqref{eq4.39},
in terms of $\hC_+,\ \hC_-,\ \bI,\ \bP,\ \bK$:
\begin{equation}
\label{eq4.40}
\begin{alignedat}{2}
&\proj_1'=\frac{1}{2}(\bI-\bP)+2\hC_- \equiv \proj_1\Big|_{M=8},
&\qquad &\dim=350, \\
&\proj_2'=-2\hC_- \equiv  \proj_2\big|_{M=8},
&\qquad &\dim=28, \\
&\proj_3'=\frac{1}{28}\bK \equiv  \proj_3\Big|_{M=8},
&\qquad &\dim=1, \\
&\proj_4'=\frac{1}{6}(\bI+\bP)-2\hC_+-\frac{1}{12}\bK
\equiv  (\proj_5 + \proj_6) \Big|_{M=8},
&\qquad  &\dim=105,
\\
&\proj_5'=\frac{1}{3}(\bI+\bP)+2\hC_++\frac{1}{21}\bK
\equiv  \proj_4\Big|_{M=8},
&\qquad&\dim=300,
\end{alignedat}
\end{equation}
where $\proj_k' \equiv \proj_{a_k}'$,
the projectors $\proj_i$ are defined in~\eqref{eq4.28} and  written for $M=8$ with the help of the identity \eqref{eq4.12}. The right hand side of \eqref{eq4.40} contains the dimensions of the eigenspaces calculated by the formulae \eqref{eq4.31}.

It is known that $\ad^{\otimes 2}(so(8))$ is decomposed
 into irreducible representations as follows (see e.g. Table 17.2 in \cite{23}, and
\cite{21}):
\begin{equation}
\label{eq4.41}
\ad^{\otimes 2}(so(8))=1+28+35+35'+35''+300+350\; .
\end{equation}
The subspaces, on which each of the irreducible representations from ~\eqref{eq4.41} act, will be denoted by $V_1$, $V_{28}$, $V_{35}$ etc. Comparing the dimensions of the subrepresentations in~\eqref{eq4.40} and \eqref{eq4.41} yields that $\proj_4^\prime$ projects~$V_{\ad}^{\otimes 2}$ onto $V_{35}+V_{35'}+V_{35''}$.
Note that for the projector $\proj_4^\prime$
the following decomposition holds:
\begin{equation}
\label{eq4.42}
\proj_4'=\proj_5|^{}_{M=8}
+\proj_6|^{}_{M=8},
\end{equation}
where $\proj_5$
and $\proj_6$ are defined in~\eqref{eq4.29} and \eqref{eq4.30}.
Besides, according to \eqref{eq4.32} the dimension of the subspace
$V_{a_5}=\proj_5|^{}_{M=8} (V_{\ad} \otimes V_{\ad})$
is $70$, while the dimension of the subspace
$V_{a_6}=\proj_6|^{}_{M=8} (V_{\ad} \otimes V_{\ad})$
equals~$35$. Hence, $V_{a_6}$~is the space of a $35$-dimensional irreducible representation of the algebra $so(8)$ (say, $V_{35''}$), and the space
$V_{a_5}$, acquired by the action of the complete antisymmetrizer\eqref{eq4.29},
must decompose into the sum $V_{35}+V_{35'}$ of the two remaining $35$-dimensional irreducible representations.

In order to build the projectors onto the invariant subspaces $V_{35}$ and $V_{35'}$
we introduce an operator $E_8\colon V_{8}^{\otimes 4}\to V_{8}^{\otimes 4}$ with the following components:
\begin{equation}
(E_8)^{i_i\dots i_4}{}_{j_1\dots j_4}
 \equiv \frac{1}{4!}\,\varepsilon^{i_i\dots i_4}{}_{j_1\dots j_4}=\frac{1}{4!}\,
 c_{j_1i_5} \cdots c_{j_1i_8}\varepsilon^{i_1\dots i_8}.
\end{equation}
Here $\varepsilon^{i_1\dots i_8}$~is the completely antisymmetric invariant tensor, $\varepsilon^{12345678}=1$. This operator, being built from the invariant tensors $\varepsilon^{i_1\dots i_8}$ and $c_{i_1i_2}$, is $\ad$-invariant. Taking into account, that in the case of algebra $so(8)$ we have $c_{i_1i_2}=\delta_{i_1i_2}$, we do not distinguish between upper and lower indices.

To build the projectors onto the eigenspaces of the operator $E_8$ we firstly find the characteristic identity for this operator.
It is convenient to perform all the calculations for a more general case of the algebra $so(2r)$, that is, for the $(2r)^r$-dimensional space $V_{2r}^{\otimes r}$ and the completely antisymmetric tensor $\varepsilon$
of the rank $2r$. The operator $E_r$ in this case is defined as:
\begin{equation}
\label{eq4.44}
(E_r)^{i_i\dots i_r}{}_{j_1\dots j_r} =\frac{(-1)^{{r}/{2}}}{r!}\,
\varepsilon^{i_i\dots i_r}{}_{j_1\dots j_r}.
\end{equation}
Note, that the following relations hold:
\begin{equation}
\label{eq4.45}
\left.
\begin{array}{l}
A_r \, E_r =E_r \, A_r =E_r\\
E_r^2=A_r
\end{array}\right\}
\implies E_r^3=E_r,
\end{equation}
where $A_r$~is the complete antisymmetrizer in the space
of all rank $r$ tensors. The components of $A_r$ are:
\begin{equation}
\label{eq4.46}
(A_r)^{i_1\dots i_r}_{\qquad k_1\dots k_r}=
\frac{1}{r!}\sum_{\sigma \in S_r} (-1)^{p(\sigma)}\delta^{i_{\sigma(1)}}_{k_1}\dots\delta^{i_{\sigma(r)}}_{k_r}
=\frac{1}{(r!)^2}\varepsilon^{i_1\dots i_rj_1\dots j_r}\varepsilon_{k_1\dots k_rj_1\dots j_r}.
\end{equation}
Here the summation is carried over all the permutations $\sigma \in S_r$,
$p(\sigma)$ is the parity of a permutation $\sigma$.
The characteristic identity \eqref{eq4.45} for $E_r$ can be written in the following way:
\begin{equation}
\label{eq4.47}
E_r (E_r -\obI\,)(E_r +\obI\,)=0,
\end{equation}
where $\obI$ denotes the identity operator in~ $V_{2r}^{\otimes r}$.

Using \eqref{eq4.47}, we can build the projectors $\oproj_1$, $\oproj_2$ and $\oproj_3$ onto the eigenspaces of $E_r$:
\begin{equation}
\oproj_{(b_i)}\equiv\oproj_j=\prod_{\substack{i=1,\\i\neq j\hfill}}^3
\frac{E_r -b_i\obI}{b_j-b_i},
\end{equation}
where $b_1=0$, $b_2=1$ and $b_3=(-1)$ are the eigenvalues of $E_r$. The exact formulae are:
\begin{equation}
\label{eq4.49}
\begin{aligned}
&\oproj_1=-(E_r-\obI\,)(E_r +\obI\,)=\obI-A_r,\\
&\oproj_2=\frac{1}{2}\,E_r (E_r +\obI\,)=\frac{1}{2}(A_r +E_r ),\\
&\oproj_3=\frac{1}{2}\,E_r (E_r -\obI\,)=\frac{1}{2}(A_r -E_r ).
\end{aligned}
\end{equation}
Here we used the second formula in the curly brackets in \eqref{eq4.45} to simplify the expressions.

 According to \eqref{eq4.49} we have
 $\oproj_2+\oproj_3=A_r$, and the images of the operators $\oproj_2$ and $\oproj_3$
 belong to the space $V_{2r}^{\wedge r}$
 of completely antisymmetric tensors of rank $r$.
 The operators $\oproj_2$ and $\oproj_3$ split the space~$V_{2r}^{\wedge r}$
 into two parts, which are called the spaces of self-dual and anti-self-dual tensors. The operator $\oproj_1=\obI-A_r $ acts trivially on the space $V_{2r}^{\wedge r}$, thus, we will not use the projector $\oproj_1$ in what follows.

In order to calculate the traces of the second and the third projectors in~\eqref{eq4.49} and to find the dimension of the corresponding subspaces we will use the following auxiliary traces:
\begin{equation}
\label{eq4.50}
\tr A_r =\binom{2r}{r}=\frac{(2r)!}{(r!)^2},\qquad
\tr E_r =0.
\end{equation}
As a result, the traces of the projectors $\oproj_2$ and $\oproj_3$ have the form:
\begin{equation}
\label{eq4.51}
\tr\oproj_2=\frac{1}{2}\,\frac{(2r)!}{(r!)^2},\qquad
\tr\oproj_3=\frac{1}{2}\,\frac{(2r)!}{(r!)^2}.
\end{equation}

Let us return now to the algebra $so(8)$ and substitute the value $r=4$ into the expressions for the traces \eqref{eq4.51}:
\begin{equation}
\label{eq4.52}
\tr\oproj_2 |^{}_{r=4}=35,\qquad
\tr\oproj_3 |^{}_{r=4}=35
\end{equation}
(hereinafter we will use $\oproj_i$ instead of $\oproj_i|^{}_{r=4}$
for simplicity).
Thus, the dimensions of the subspaces $\oproj_2 \cdot V_{8}^{\wedge 4}$ and $\oproj_3 \cdot V_{8}^{\wedge 4}$ coincide and equal~$35$. Recall that the projector $\proj_5$, given in~\eqref{eq4.29}, for every $M$ equals the complete antisymmetrizer $A_4$, and, in particular $\proj_5|^{}_{M=8}=A_4$, hence
\begin{align}
\proj_5|^{}_{M=8}=\oproj_2+\oproj_3,
\label{eq4.53}
\end{align}
and the images of $\oproj_2$ and $\oproj_3$
belong to the subspace extracted by the projector $\proj_5|^{}_{M=8}=A_4$.
The values of the traces \eqref{eq4.52}
suggest that $\oproj_2$ and $\oproj_3$ are the projectors onto the two $35$-dimensional invariant subspaces of the representation $\ad^{\otimes 2}(so(8))$.

To write the projector $\proj_6|^{}_{M=8}$ in terms of the operators $E_4$, $A_4$, introduced in~\eqref{eq4.44} and \eqref{eq4.46}, and in terms of the operators $\bI$, $\bP$, $\bK$ and $\hC_+$ we use the formula \eqref{eq4.42}, the explicit form of the projector $\proj_4'$, given in~ \eqref{eq4.40}, as well as the fact that
$\proj_5|^{}_{M=8}=A_4$. As a result, we have:
\begin{equation}
\proj_6|^{}_{M=8}=\proj_4'-\proj_5|^{}_{M=8}
=\frac{1}{6}(\bI+\bP)-2\hC_+-\frac{1}{12}\,\bK-A_4.
\end{equation}
Finally, we have the full system of mutually orthogonal and primitive projectors onto the spaces of the irreducible subrepresentations in~$\ad^2(so(8))$:
\begin{equation}
\begin{alignedat}{2}
&\proj_1^\prime=\frac{1}{2}(\bI-\bP)+2\hC_-, &\qquad &\dim=350,\\
&\proj_2^\prime=-2\hC_-, &\qquad &\dim=28,\\
&\proj_3^\prime=\frac{1}{28}\bK, &\qquad &\dim=1,\\[1mm]
&\proj_6|^{}_{M=8}=\frac{1}{6}(\bI+\bP)-2\hC_+-\frac{1}{12}\,\bK-A_4,
&\qquad &\dim=35,\\[1mm]
&\oproj_2=\frac{1}{2}(A_4+E_4), &\qquad &\dim=35,\\[1mm]
&\oproj_3=\frac{1}{2}(A_4-E_4), &\qquad &\dim=35,\\[1mm]
&\proj_5^\prime=\frac{1}{3}(\bI+\bP)+2\hC_++\frac{1}{21}\,\bK, &\qquad &\dim=300.
\end{alignedat}
\end{equation}

\section{A connection between the eigenvalues of the operator $\hC$ in the adjoint representation of $so(N,\mathbb{C})$ and $sp(N,\mathbb{C})$ Lie algebras and the Vogel parameters}
\label{s5}
Many results of this paper, namely, construction of
the characteristic identities and
projectors onto invariant subspaces of the representation $\ad^{\otimes 2}(\mfg_N^\epsilon)$ in terms of the split Casimir operator, as well as calculation of the dimensions of these subspaces, could be also achieved by using the so-called Vogel parameters \cite{14},~\cite{18}. These parameters are defined as three numbers $(\alpha,\beta,\gamma)$ modulo a common multiplier and an arbitrary permutation from the symmetric group $S_3$ (and thus can be interpreted as coordinates in the space $\mathbb{P}^2/S_3$). It is known \cite{14}, \cite{15}, \cite{18} that certain values of the Vogel parameters (or what is the same, a certain point in the space $\mathbb{P}^2/S_3$) correspond to some
simple Lie algebra. For the simple Lie algebras of classical series values of these parameters are given in Table~\ref{tab3}~
(see \cite{17}), where $t\equiv \alpha+\beta+\gamma$~is the dual Coxeter number.

\begin{table}[h!]

\small
\centering
\caption{Vogel parameters for simple Lie algebras.}
\label{tab3}

\vspace*{1mm}
\tabcolsep=1.5em
\renewcommand{\arraystretch}{1.2}

\begin{tabular}{|c|c|c|c|c|c|}
\hline
Type & Lie Algebra & $\alpha$ & $\beta$ & $\gamma$ & $t$\\
\hline
$A_r$ & $sl(r+1)$ & $-2$ & $2$ & $r+1$ & $r+1$\\
\hline
$B_r$ & $so(2r+1)$ & $-2$ & $4$ & $2r-3$ & $2r-1$\\
\hline
$C_r$ & $sp(2r)$ & $-2$ & $1$ & $r+2$ & $r+1$\\
\hline
$D_r$ & $so(2r)$ & $-2$ & $4$ & $2r-4$ & $2r-2$\\
\hline
\end{tabular}
\end{table}

\noindent
The representation $\ad^{\otimes 2}\mcA$ of
any simple Lie algebra ~$\mcA$ can be decomposed into symmetric and antisymmetric subspaces. According to papers
 \cite{23}, \cite{14}--\cite{16} the symmetric part (in all the cases of
 simple Lie algebras of classical series, apart from $sl(2,\mathbb{C})\cong sp(2,\mathbb{C})\cong so(3,\mathbb{C})$, $sl(3,\mathbb{C})$ and $so(8,\mathbb{C})$), in its turn, decomposes into four irreducible subrepresentations -- a singlet $T_0$ and three irreducible representations denoted by $Y_2^{(\alpha)}$, $Y_2^{(\beta)}$ and $Y_2^{(\gamma)}$, where $\alpha$, $\beta$ and $\gamma$~are the Vogel parameters of the algebra $\mcA$ from Table~\ref{tab3}. The dimensions of these four representations as well as the eigenvalues of the quadratic Casimir operator $C_{(2)}$ (which was defined in~\eqref{eq3.5}), acting on the spaces of these representations, are given by formulas \cite{14},~\cite{18}
\begin{equation}
\label{eq5.1}
\begin{alignedat}{2}
&\dim T_0(\mcA)=1, &\qquad &c_2^0=0,\\
&\dim Y_2^{(\alpha)}(\mcA)=-\frac{(3\alpha-2t)(\beta-2t)(\gamma-2t)t(\beta+t)(\gamma+t)}{\alpha^2(\alpha-\beta)\beta(\alpha-\gamma)\gamma}, &\qquad &c_2^\alpha=2-\frac{\alpha}{t}\,,\\
&\dim Y_2^{(\beta)}(\mcA)=-\frac{(3\beta-2t)(\alpha-2t)(\gamma-2t)t(\alpha+t)(\gamma+t)}{\beta^2(\beta-\alpha)\alpha(\beta-\gamma)\gamma}, &\qquad &c_2^\beta=2-\frac{\beta}{t}\,,\\
&\dim Y_2^{(\gamma)}(\mcA)=-\frac{(3\gamma-2t)(\beta-2t)(\alpha-2t)t(\beta+t)(\alpha+t)}{\gamma^2(\gamma-\beta)(\beta(\gamma-\alpha)\alpha}, &\qquad &c_2^\gamma=2-\frac{\gamma}{t}\,,
\end{alignedat}
\end{equation}
where $c_2^0$, $c_2^\alpha$, $c_2^\beta$ and $c_2^\gamma$~are the eigenvalues of the operator $C_{(2)}$, acting in the representations $T_0$, $Y_2^{(\alpha)}$, $Y_2^{(\beta)}$, $Y_2^{(\gamma)}$ respectively (here the upper index of $c_2$ is not the highest weight, as in the formula \eqref{eq3.10}, but is chosen to accord with the notation of the corresponding representation).

The antisymmetric part decomposes into the direct sum of two representations, one of which is the adjoint representation $\ad$. The dimension of this representation and the value $c_2^{\ad}$ of the quadratic Casimir operator $C_{(2)}$ in it are expressed by the following formulae:
\begin{equation}
\label{eq5.2}
\dim (\ad)\equiv \dim \mcA=\frac{(\alpha-2t)(\beta-2t)(\gamma-2t)}{\alpha\beta\gamma}, \qquad c_2^{\ad}=1.
\end{equation}
The second antisymmetric subrepresentation is denoted $X_2$. Its dimension and the value $c_2^X$ of the operator $C_{(2)}$ in it are:
\begin{equation}
\label{eq5.3}
\dim (X_2)=\frac{1}{2}\dim \mcA(\dim\mcA-3), \qquad c_2^X=2.
\end{equation}
 Comparing the dimensions \cite{21} of the irreducible representations in~$\ad^{\otimes 2}$ with the dimensions \eqref{eq4.32}
 indicates that
 the representation~$X_2$ for the Lie algebras of the types $B_r,C_r$
 and~$D_r$ is irreducible. Note that the representation~$X_2$ for Lie algebras of the type $A_r$, $r>1$, is reducible and decomposes into the sum of two irreducible representations with the same dimensions.

The final decomposition of the representation $\ad^{\otimes 2}(\mcA)$,
where $\mcA$~is a classical simple Lie algebra (except for the algebras
$sl(2,\mathbb{C})\cong sp(2,\mathbb{C})\cong so(3,\mathbb{C})$,
$sl(3,\mathbb{C})$ and~$so(8,\mathbb{C})$), takes the form
\begin{equation}
\label{eq5.4}
\ad^{\otimes 2}(\mcA)=T_0(\mcA)+Y_2^{(\alpha)}(\mcA)
+Y_2^{(\beta)}(\mcA)+Y_2^{(\gamma)}(\mcA)+\ad(\mcA)+X_2(\mcA).
\end{equation}

If we know the values \eqref{eq5.1}--\eqref{eq5.3}
of the quadratic Casimir operator $C_{(2)}$ in the six representations
(in the case of those representations being irreducible), we can calculate the values of the split Casimir operator $\hC$ (which was defined in~\eqref{eq3.3}) in these representations by using the formula \eqref{eq3.9}. A connection between the eigenvalues of $\hC$ and $C_{(2)}$ in any irreducible subrepresentation
 $T_{\lambda}$ in the decomposition $\ad\otimes\ad$ with the help of the formula \eqref{eq3.9}, where one needs to set the values
  $\lambda_1$ and $\lambda_2$ equal to the highest weight of the irreducible representation of the algebra $\mcA$. As the values of the quadratic Casimir operator of any simple lie Algebra $\mcA$ in the adjoint representation equal $1$, (see e.g. \cite{20}), then $c_2^{(\lambda_1)}=c_2^{(\lambda_2)}=1$ and according to
 \eqref{eq3.9} we have
\begin{equation}
\label{eq5.5}
\hc^{(\lambda)}=\frac{1}{2}c_2^{(\lambda)}-1.
\end{equation}

In what follows we will only consider the cases of the algebras $so(N,\mathbb{C})$ and $sp(N,\mathbb{C})$, which our paper is devoted to. Substituting the values $\alpha$, $\beta$,
 $\gamma$ and $t$ for the algebras $so(N,\mathbb{C})$ and $sp(N,\mathbb{C})$ from Table~\ref{tab3} into the formulas \eqref{eq5.1}--\eqref{eq5.3}, yields that the dimensions of the six representations $T_0$, $Y_2^{(\alpha)}$, $Y_2^{(\beta)}$, $Y_2^{(\gamma)}$, $\ad(\mfg_N^\epsilon)$ and $X_2$ coincide with the dimensions \eqref{eq4.32} of the invariant subspaces of the representations $\ad^{\otimes 2}(\mfg_N^\epsilon)$.
 Taking into account the formula \eqref{eq5.5} one can show, that the values of the split Casimir operator in each of the aforementioned representations coincide with the roots \eqref{eq4.22} of the polynomial on the left hand side of~\eqref{eq4.26}, which is the characteristic polynomial of $\hC$ in the representation $\ad^{\otimes 2}(\mfg_N^\epsilon)$. It allows us to conclude that, the spaces of the representations on the right hand side of~\eqref{eq5.4}
coincide with the invariant subspaces, that were extracted by the projectors~\eqref{eq4.28}.This correspondence is shown in Table \ref{tab4}, where in the top row all the six projectors \eqref{eq4.28} are given, while in the second, third and forth rows there are subrepresentations, subspaces of which are extracted in~$V_{\ad} \otimes V_{\ad}$ by the projectors in the top row.
\begin{table}[h!]
\centering

\tabcolsep=1.5em
\caption{Correspondence between the representations and the projectors.}
\label{tab4}

\renewcommand{\arraystretch}{1.2}

\vspace*{1mm}
\begin{tabular}{|c|c|c|c|c|c|c|}
\hline
& $\proj_1$ & $\proj_2$ & $\proj_3$ & $\proj_4$ & $\proj_5$ & $\proj_6$\\
\hline
$B_r$ & $X_2$ & $\ad(so(2r+1))$ & $T_0$ & $Y_2^{(\alpha)}$ & $Y_2^{(\beta)}$ & $Y_2^{(\gamma)}$\\
\hline
$C_r$ & $X_2$ & $\ad(sp(2r))$ & $T_0$ & $Y_2^{(\beta)}$ & $Y_2^{(\alpha)}$ & $Y_2^{(\gamma)}$\\
\hline
$D_r$ & $X_2$ & $\ad(so(2r))$ & $T_0$ & $Y_2^{(\alpha)}$ & $Y_2^{(\beta)}$ & $Y_2^{(\gamma)}$\\
\hline
\end{tabular}
\end{table}

Thus, we obtain that the calculated in the previous sections values of the split Casimir operator in the irreducible subrepresentations in~$\ad^{\otimes 2}$ for the cases of $so(N,\mathbb{C})$ and~$sp(N,\mathbb{C})$ Lie algebras, as well as the dimensions of the corresponding invariant subspaces could be written in terms of the Vogel parameters in complete accordance with papers \cite{14}--\cite{16}.

\section{Conclusion}
\label{s6}
In our work we obtained explicit formulae for
projectors onto the spaces of the irreducible subrepresentations comprising the tensor product of two adjoint representation for all complex simple Lie algebras $so(N,\mathbb{C})$ for $N\ge 3$ and $sp(N,\mathbb{C})$ for $N\ge 2$. For Lie algebras $so(N,\mathbb{C})$ ($N\ge 3$)
the analogous projectors
where obtained in \cite{23} (see Table 10.3)
by using the diagram technique. We proved the natural conjecture
 that the projectors for algebras $sp(N,\mathbb{C})$ $(N=2r)$
 are related to the corresponding projectors for algebras $so(N,\mathbb{C})$ by the duality
 transformation\footnote{Generally speaking this duality
 does not work for representations which can not be constructed
 from the adjoint representations.}
  $N \to - N$.
 In all considered cases, except the case of
 algebra $so(8)$, the formulation was conducted by finding the characteristic identities for the split Casimir operator $\hC$
 in adjoint representation. In  special case of $so(8,\mathbb{C})$ (considered separately) an additional invariant operator
 (independent of $\hC$) was used to construct the complete set of projectors. This operator allows to decompose the completely
 antisymmetric representation of rank $4$ into
 the selfdual and the anti-selfdual parts.
 It was then shown that the dimensions of the irreducible subrepresentations in $\ad^{\otimes 2}$,
 as well as the values of the quadratic Casimir opertor in these subrepresentations are in complete correspondence with the results of papers \cite{14}--\cite{16}, where these values were written in terms of the Vogel parameters.

\subsection*{Acknowledgements}
The authors would like to thank P.~Cvitanovi\'c, S.O.~Krivonos,
 and O.V.~Ogievetsky for useful discussions. The work of  A.P.~Isaev was supported by the Russian Foundation for Basic Research grant 19-01-00726. The work of A.A.~Provorov was supported by the Russian Foundation for Basic Research grant 20-52-12003\textbackslash 20.

\end{document}